\documentclass[a4paper,twocolumn]{article}

\usepackage{amsmath}
\usepackage{graphicx}
\usepackage[noblocks]{authblk}
\usepackage[top=3cm,bottom=2cm,left=2cm,right=2cm]{geometry}
\usepackage{flushend}
\usepackage{multirow}

\begin{document}

\title{\bf BINODAL LAYER IN ISENTROPICALLY EXPANDING SLAB TARGET WITH VAN DER
WAALS EQUATION OF STATE}
\author{Borovikov\,D.}
\affil{MIPT, Dolgoprudny, Russia}
\author{Iosilevskiy\,I.}
\affil{JIHT RAS, Moscow, Russia}
\date{}
\maketitle

\textbf{Introduction.} Features of isentropic expansion of warm dense matter (WDM) created by intense energy fluxes (strong shock compression or instant isochoric heating by laser or heavy ion beams) are under discussion in situation when (i) – thermodynamic trajectory of such expansion crosses binodal of liquid-gas phase transition, and (ii) – expansion within two-phase region is going along equilibrium (not metastable) branch of the isentrope for two-phase mixture. It is known for the case of plane geometry \cite{Anisimov} \cite{Inogamov} that because of sharp break of the expansion isoentrope at boiling point (in P−V plane) there appears extended zone (liquid layer) of uniformity for expanding material with constant thermodynamic and kinematic parameters, which correspond exactly to the state on binodal. General properties of such boiling (’binodal’) layer were claimed and discussed at \cite{ILIos} for the cases of isentropic expansion of infinite sample as well as for finite plane and spherical samples (slab and ball) and for the system of well positioned slabs (stuck target).

\textbf{Origin of boiling layer.} It is known \cite{Zeldovich} that isoentropic expansion of semi-infinite layer along isoentrope with negative break in P–V plane (decreasing of sound speed) is accompaied with formation of a uniform layer of matter (’plateau’) with constant therodynamic and kinematic parameters. In particular in \cite{Anisimov} formation of such liquid layer was studied in the case of such break of isoentrope at boiling boundary of gas-liquid phase transition. In \cite{ILIos} this layer was reffered to as Boiling
(or ’Binodal’) Layer (BL). In \cite{borovik} formation of a uniform binodal layer was studied thoroughly for isoentropic expansion of semi-infinite layer of Van derWaals (VdW) fluid. Goal of the work was to prove theoretically and numerically the phenomenon and initiate further investigation. VdW equation of state (EOS) is chosen because it allows in the simplest planar case to combine both analytical and numerical methods to obtain very accurate results.

Present work is aimed at studying behavior of vdW fluid in the case of isoentropic expansion of finite layer.

\textbf{Self-similarity.} Consider finite layer of vdW fluid with uniform initial parameters, no mass velocity and thickness $2\Delta$. It is convenient to introduce such coordinate system that the middle point of the layer is placed in the $x=0$. In this case solution is either odd (thermodynamical parameters) or even (velocity) function of $x$. So further (unless it is stated otherwise) only "positive" ($x>0$) part of it will be reffered to.

Reduced vdW EOSs in one- and two-phase regions are respectively \cite{Landau5}, \cite{Kraikoetal}:
\begin{align}
\label{EOS}
e(p,r)&=(p +  3r^{2})\frac{3 - r}{2r}-3r \nonumber \\
e(p,r)&=\frac{f(p)}{r}+g(p)
\end{align}

Governing equations of the process are hydrodynamical equations closed by EOS \cite{Landau6}, \cite{Kraiko}:
\begin{align}
&\frac{\partial r}{\partial t}+\frac{\partial r u}{\partial x}=0;\nonumber\\
&\frac{\partial u}{\partial t}+u\frac{\partial u}{\partial  x}=-\frac{1}{r}\frac{\partial p}{\partial x};\nonumber\\
&\frac{\partial e}{\partial t}+u\frac{\partial e}{\partial x}=-\frac{p}{r}\frac{\partial u}{\partial x};\nonumber\\
&e=e(p,r).\label{HydEq}
\end{align}
All quantities in the system \eqref{HydEq} are dimensionalized with critical parameters used as characteristic scale for pressure, density and energy. For space variable scale $\Delta$ is used, characteristic value of the velocity is $\sqrt{P_{crit}/\rho_{crit}}$. All quantities further are dimensionless.

In the planar 1D case solution is self-similar: all quantitie depend on variable $\xi=x/t$ only \cite{Landau6, borovik}. In the case considered in the present work analagous situation takes place in the beginning of the expansion: before disturbance reaches the middle point of the layer, solution can be represented as combination of two independent self-similar functions.

Adiabatic curve should also have different expressions in one- and two-phase regions. The former can be derived analytically:
\begin{displaymath}
p_{s}(r)=\text{const}\left(\frac{r}{3-r}\right)^{\gamma_{0}}-3r^{2}.
\end{displaymath}

In two-phase region binodal curve can be calculated numerically as a solution of ordinary differential equation ($v=1/r$):
\begin{displaymath}
\left(\frac{\partial p}{\partial v}\right)_{s}=-\frac{\left(\frac{\partial e}{\partial v}\right)_p+p}{\left(\frac{\partial e}{\partial p}\right)_v}=-\frac{f(p)+p}{f'(p)v+g'(p)}.
\end{displaymath}

\begin{figure}[h]
    \includegraphics[width=0.99\columnwidth]{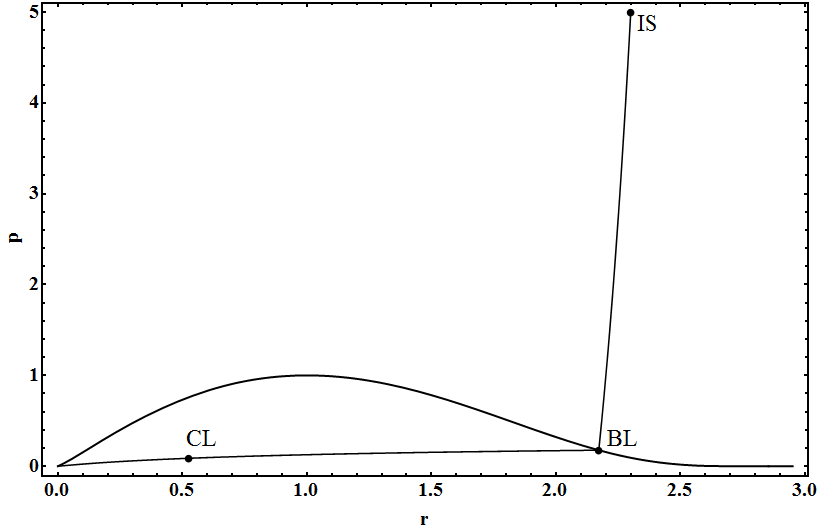}
    \caption{Phase diagram. IS - initial state, BL - boiling layer, CL - core layer.}    
    \label{phasediag1}
\end{figure}

Further description consideres phenomenon with initial parameters $r_{0}=2.3$, $p_{0}=5$. Phase diagram with Poisson adiabat is shown on \mbox{Figure \ref{phasediag1}}.

At the conclusion of this section it should be noted that in dimensionalized equations \eqref{HydEq} parameter $\Delta$ falls out completely. This means that no dimensionless parameter (like Reynolds number) appears in the equations and phenomenon is not qualitatively affected by value of $\Delta$. For fixed dimensionless initial parameters of matter time of the process is scaled as $\Delta \sqrt{\rho_{crit}/P_{crit}}$.

\begin{figure}[h]
    \includegraphics[width=0.99\columnwidth]{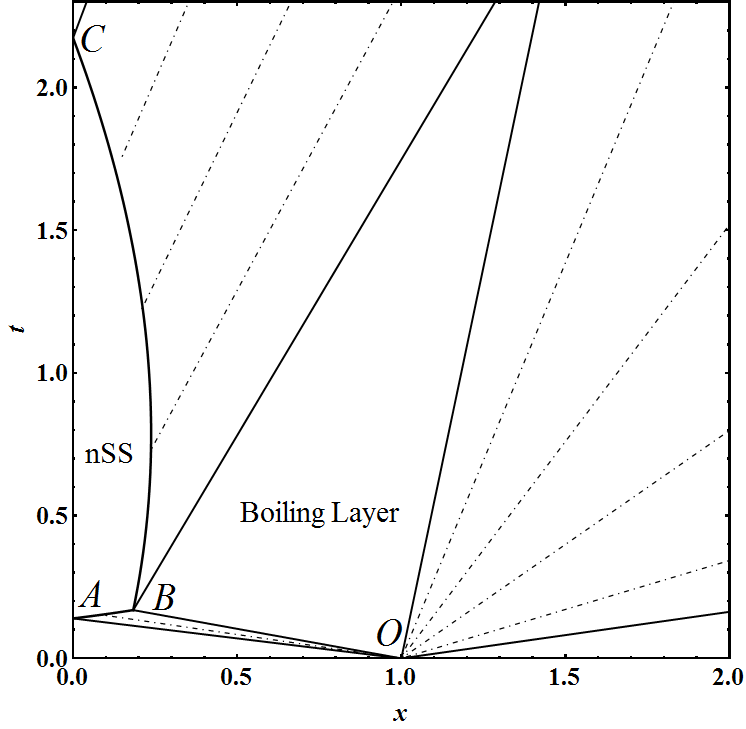}
    \caption{$x-t$ diagram of the beginning of expansion of the layer.}    
    \label{XTdiag1}
\end{figure}

\textbf{Non self-similarity.} As soon as disturbance reaches the middle point of the layer (point $A$ on \mbox{Figure \ref{XTdiag1}}) the solution cannot be considered as self-similar. This means that semi-analytical approached used in \cite{borovik} is no longer applicable (at least in the initial form). However velocity still can be calculated as:
\begin{equation}
\label{u_change}
u=u_{0} \pm \bigg| \int_{r_{0}}^{r} c\; \frac{\mathrm{d}r'}{r'} \bigg|,
\end{equation}
where sign before integral is determined by whether fluid is accelerated or decelerated, integral is taken along a characteristic. 

Also the process is still isoentropic so all states of a fluid belong to the same curve $p_{s}(r)$.

\textbf{Method of characteristics}. Though exact solution cannot be obtained, qualitative analysis with several estimations such as times and locations of major events in the process can be performed. Characteristics should be considered: their positions on $x-t$ plane describe all the features of the expansion.

At any moment $t_{0}$ and location $x_{0}$ slope of the characteristic curve is \cite{Landau6}:
\begin{equation}
\label{zeta}
\zeta_{\pm}(x_{0},t_{0})=\frac{\mathrm{d} x}{\mathrm{d} t} \bigg|_{\zeta_{\pm}}= u(x_{0},t_{0}) \pm c(x_{0},t_{0}).
\end{equation}

In self-similar case these curves are straight lines on $x-t$ plane: family $\zeta_{-}$ originates in the point $(1,0)$ (point $O$ on  \mbox{Figure \ref{XTdiag1}}), family $\zeta_{+}$ - in the point $(-1,0)$.

Since speed of sound of vdW fluid is discontinuous across binodal, all characteristics in each family can be divided into two sets separated by region associated with BL, which is clearly seen on \mbox{Figure \ref{XTdiag1}}.

Thick lines on \mbox{Figure \ref{XTdiag1}} are charactristic curves that play important role such as being border of non self-similarity (nSS) region (like curves $AB$ and $BC$) or bounds of characteristics' sets (like $OA$, $OB$ etc.). Dashed lines are characteristic curves within family.

Physical meaning of characteristic is propagation of information across the matter \cite{Landau6}. So the set confined between lines $OA$ and $OB$ is rarefaction of fluid in the original phase while the other one is rarefaction of fluid in two-phase state. Region that separates these two sets contains no characteristics so the matter inside it stays undisturbed and uniform. This is BL which corresponds to state exactly on binodal.

Before disturbace reaches point $A$ the whole solution remains self-similar. The moment of meeting of two families is:
\begin{equation}
\label{tA}
t_{A}=\frac{1}{c_{0}},
\end{equation}
where $c_{0}$ is speed of sound in initial state of the fluid.

On intersection with characteristics from another family slope of any given characteristic continuously changes \cite{Landau6}. If  behavior of one family is known then another one is described by ordinary differential equation which is derived from \eqref{zeta}:
\begin{equation}
\label{zeta1}
\frac{\mathrm{d} x}{\mathrm{d} t} \bigg|_{\zeta_{\pm}}=2u(x,t)-\zeta_{\mp}(x,t).
\end{equation}

\textbf{Non self-similar region}. Relation \eqref{zeta1} allows to calculate line $AB$ ad dependence $t_{AB}(x)$: behavior of this set of characteristics is self-similar. Point B is determined by intersection of $t_{AB}(x)$ and characteristic with slope $\frac{\mathrm{d} t}{\mathrm{d} x}=(u_{b}-c_{b}^{+})^{-1}$, where $u_{b}$ is mass velocity of matter in BL, $c_{b}^{+}$ is sound of speed before entering two-phase region:
\begin{equation}
\label{tB}
t_{B}=t_{AB}(x_{B})=\frac{x_{B}-1}{u_{b}-c_{b}^{+}}
\end{equation}

Moment $t_{B}$ plays very important role in the process. One can notice that density at point $x_{B}$ is exactly $r_{b}$, which is density of matter in BL. Further interaction of colliding rarefaction waves will decrease it so\mbox{ $r(x_{B},t>t_{B})<r_{b}$}. This leads to both characteristic, that intersect at this point, having a break in the slope: as density crosses value $r_{b}$ speed of sound undergoes a jump. Therefore slope increases on $x-t$ diagram: rate of propagtion of information slows down at this point, and it occurs jump-like. If one supposes that in region $|x|<x_{B}$ there is a portion of matter in one-phase state, that would inevitably lead to intersection of characteristics from the same family. This means formation of a shock wave in the system \cite{Landau6} and hence increase in entropy. But the process of expansion into vacuum is isoentropic, so the assumption was wrong.

Conclusion is: at the moment $t_{B}$ in the whole region $x<x_{B}$ the matter has the same thermodynamic parameters as the one in BL. This is the only condition that allows all characteristics in the region to undergo slope jump and prevents them from intersections with each other.

So $t=t_{B}$ is the moment when an extensive zone of uniform thermodynamical parameters exist in the center of the system. It includes both BLs and a portion of matter between them. Also at that moment and later all matter is in two-phase region and it is described by $2^{nd}$ equation in \eqref{EOS} only.

Jump of sound of speed (and therefore of slope of characteristics) depends which point of binodal Poisson adiabat intersects. In the case shown on \mbox{Figure \ref{XTdiag1}} it is big enough to change the sign of characteristic's slope. But evetually it turns back and intersects with axis of symmetry at point $C$ (otherwise density on this axis would never approach zero). All characteristics that intersect line $BC$ leave nSS region as straight lines since they don't encounter characteristics from another set immediately. That means that they comprise self-similar rarfaction wave and behave as if they originated in some imaginary point $I$ (which is in $x<0$ region). If coordinates of this point are known, then slopes on line $BC$ and hence line $BC$ itself can be determined. Characteristic that leaves point $B$ has slope $(u_{b}+c_{b}^{-})^{-1}$ and the one that leaves point $C$ has slope $(c_{core})^{-1}$, where $c_{core}$ is sound of speed in state in point $C$. It can be calculated if one notices that on axis of symmetry mass velocity is always zero. That determines density in point $C$:
\begin{equation}
\label{core}
u_{b}-\bigg| \int_{r_{b}}^{r_{C}} c\, \frac{\mathrm{d} r'}{r'} \bigg|=0
\end{equation}
Relation \eqref{core} is derived from \eqref{u_change} with taking into account that fluid is decelerated along line $BC$.

With slopes of two border characteristics known position of imaginary point $I$ can be found iteratively. According to \eqref{zeta1} line $BC$ should be calculated as dependence $x_{BC}(t)$:
\begin{equation}
\label{BC}
\frac{\mathrm{d} x_{BC}}{\mathrm{d} t}= 2 u_{BC}\left(\frac{x_{BC}-x_{I}}{t-t_{I}}\right)-\frac{x_{BC}-x_{I}}{t-t_{I}},
\end{equation}
where $u_{BC}(\zeta)$ is obtained as solution of system \eqref{HydEqTr} with proper chosen signs. Test for correction of coordinates of point $I$ is slope of curve $BC$ at $t_{I}$: it shoud coincide with $-c_{core}$. If it exceeds this value ($\frac{\mathrm{d} x_{BC}}{\mathrm{d} t}>-c_{core}$) then it should be farther from the axis of symmetry; in opposite case - closer to the axis. After this iterative process is stopped, time $t_{C}$ is calculated as solution of:
\begin{equation}
\label{tC}
x_{BC}(t_{C})=0.
\end{equation}

\begin{figure}[h]
    \includegraphics[width=0.99\columnwidth]{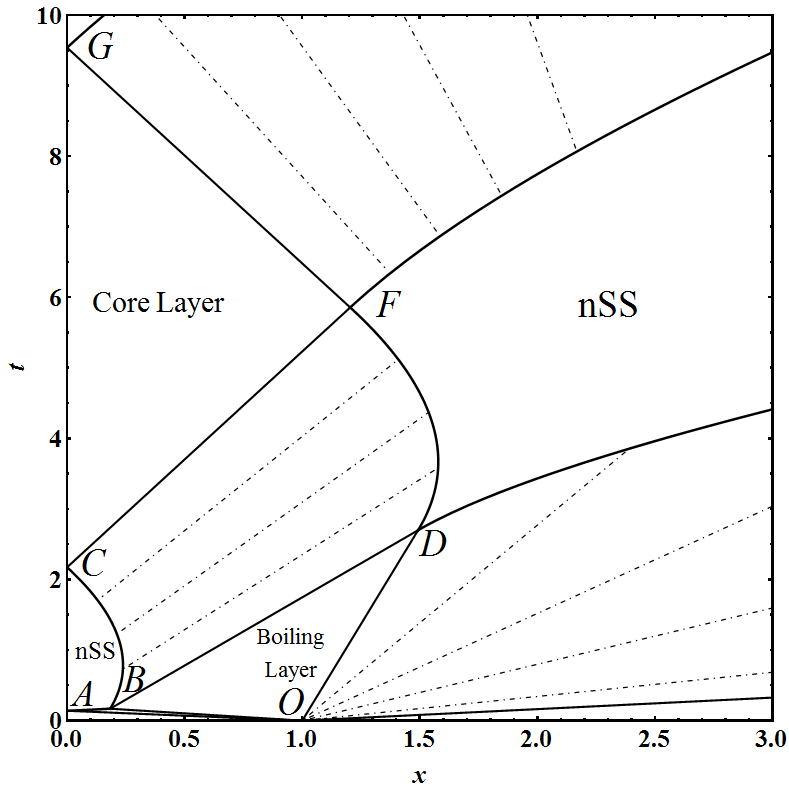}
    \caption{$x-t$ diagram of expansion of the layer.}    
    \label{XTdiag2}
\end{figure}

\textbf{Boiling and core layers.} In the simplest case consodered \mbox{in \cite{borovik}} BL exists always and its size constantly increases with rate $(c_{b}^{+}-c_{b}^{-})$. \mbox{Figure \ref{XTdiag1}} shows that in present geometry BL grows up to moment $t_{B}$ when its size reaches maximum $(c_{b}^{+}-c_{b}^{-})t_{B}$. Then it decreases with rate $2c_{b}^{-}$. BL layer completely disappear at \mbox{point D} with coordinates (\mbox{Figure \ref{XTdiag2}}): 
\begin{equation}
\label{tD}
t_{D}=\left(1+\frac{c_{b}^{+}-c_{b}^{-}}{2c_{b}^{-}}\right)t_{B} \quad x_{D}=x_{B}+2c_{b}^{-}(t_{D}-t_{B})
\end{equation}

Peculiarity of the case considered is that $t_{D}>t_{C}$, which is not true in general case. This means that during the interval $t_{C}<t<t_{D}$ solution is self-similar and can ve obtained with semi-analytical methods.

Important feature of the phenomenon is that after \mbox{moment $t_{C}$} there are no chararcteristics in the vicinity of symmetry axis. Therefore another uniform region forms (core layer - CL) and matter contained in it is at rest: mass velocity xactly equals zero. Thermodynamic parameters are the same as at the \mbox{point $C$}. CL linearly grows up to moment $t_{F}$ with rate $c_{core}$ and linearly decreases after that moment with the same rate.

Point $F$ is determined by intersection of characteristic $CF$ (that leaves point $C$ as a straight line) and curve $DF$. The latter can be calculated with the help of relation \eqref{zeta1} (as well as another charactersitic crossing point $D$ - $D\infty$):
\begin{align}
\label{DF}
\frac{\mathrm{d} x_{DF}}{\mathrm{d} t}&= 2 u_{DF}\left(\frac{x_{DF}-x_{I}}{t-t_{I}}\right)-\frac{x_{DF}-x_{I}}{t-t_{I}},\nonumber\\
\frac{\mathrm{d} x_{D\infty}}{\mathrm{d} t}&= 2 u_{D\infty}\left(\frac{x_{D\infty}-1}{t}\right)-\frac{x_{D\infty}-1}{t}.
\end{align}

Difference is that curve $DF$ should be calculated as dependence $x_{DF}(t)$, while there is no such restriction for curve $D\infty$. These two curves are part of border of another non self-similar region. Another part of the border is curve $F\infty$. It can be calculated if one notices that after leaving this nSS region charactersitics become straight lines again. Again they behave as if they originated from some imaginary point $J$.

It  should be noted here that as $x\to\infty$ both curves $D\infty$ and $F\infty$ appproach utmost characteristic. It has a slope $u_{\infty}$:
\begin{equation}
\label{u_inf}
u_{\infty}= \int_{0}^{r_{b}} c\, \frac{\mathrm{d} r'}{r'}.
\end{equation}

This utmost characteristic $O\infty$ can be considered as the limit of charactesitics that leave nSS region. So imaginary point $J$ can be found as intersection point of this limit and line $FG$ prolonged down:
\begin{equation}
\label{xJ}
\frac{x_{J}-x_{F}}{c_{core}}+\frac{x_{J}-1}{u_{\infty}}=t_{F}.
\end{equation}

With point $J$ known the whole curve $F\infty$ can be calculated with the help of relation \eqref{zeta1}:
\begin{equation}
\label{F_inf}
\frac{\mathrm{d} x_{F\infty}}{\mathrm{d} t}= 2 u_{F\infty}\left(\frac{x_{F\infty}-x_{J}}{t-t_{J}}\right)-\frac{x_{F\infty}-x_{J}}{t-t_{J}}.
\end{equation}

With coordinates of point $F$ known time when CL disappear can be calculated:
\begin{equation}
\label{tG}
t_{G}=t_{F}+\frac{x_{F}}{c_{core}}.
\end{equation}

After this moment solution is comprised by two nSS regions (one is confined between $D\infty$ and $F\infty$, the other one appear on symmetry axis) and self-similar region confined between $F\infty$ and $G\infty$.

As it was stated before, this analysis doesn't give the whole solution: parameters within nSS regions cannot be obtained. But times and location of all major events can be calculated and used as a test for any numerical method used to simulate the problem. For initial parameters considered in this work they are given in the \mbox{Table \ref{XTparams}} (accuracy is $\pm0.01$).
\begin{table}[!h]
\caption{x - t parameters}
	\begin{center}
		\label{XTparams}
		\begin{tabular}{|c|c|c|c|c|c|c|c|}
		\hline
		 & O & A & B & C & D & F & G \\
		\hline
		$x$ & 1.00 & 0.00 & 0.18 & 0.00 & 1.49 & 1.20 & 0.00 \\
		\hline
		$t$ & 0.00 & 0.14 & 0.17 & 2.17 &  2.70 & 5.86 & 9.54 \\
		\hline
		\end{tabular}
	\end{center}
\end{table}

\textbf{Acknowledgements.}

This work was supported by the RAS Scientific Program "Physics of extreme states of matter". Authors acknowledge Alexander Kraiko and Haris Valiev for fruitful discussions and valuable comments.


\begin{thebibliography}{99}

\bibitem{ILIos}
Iosilevskiy I.L. ''Phase freezeout" in isentropically expanding matter, in "Physics of Extreme States of Matter", Eds. V. Fortov et al. (Chernogolovka: IPCP
RAS), Russia, 2011, PP. 99-102
\bibitem{borovik}
Borovikov D., Iosilevskiy I. Semi-analytical calculations for parameters of boiling layer in isentropic expansion of warm dense matter with van der Waals equation of state, in "Physics of Extreme States of Matter", Eds.V. Fortov et al. (Chernogolovka: IPCP RAS), Russia, 2012, PP.12-15; (arxiv:1209.0398)
\bibitem{Landau5}
Landau L.D., Lifshitz E.M. Statistical Physics. part I.
\bibitem{Landau6}
Landau L.D., Lifshitz E.M. Hydrodynamics.
\bibitem{Zeldovich}
Zeldovich Ya., Raizer Yu., Physics of shock waves and high-T hydrodynamics, (Fizmatlit, Moscow, 2008)
\bibitem{Anisimov}
Anisimov S.I., Inogamov N.A., Oparin A.M., Rethfeld B., Ogawa M., Fortov V.E., Appl. Phys. A 69, 617 (1999)
\bibitem{Inogamov}
Inogamov N.A., Anisimov S.I., Rethfeld B., JETF 115, 2091 (1999)
\bibitem{Kraikoetal}
Iosilevskiy I.L., Borovikov D. S., Valiev H. F., Kraiko A.N., Binodal layer in isentropically expanding matter, in Proceedings of Khariton Scientific Readings, Sarov, Russia, March 2013
\bibitem{Kraiko}
Kraiko A.N., Theoretical Hydrodynamics, Torus Press, Moscow, 2010, pp.430

\end{thebibliography}
\end{document}